\documentclass[11pt]{article}

\usepackage{latexsym}

\def\beq{\begin{eqnarray}}
\def\eeq{\end{eqnarray}}
\def\ln{\,\mbox{ln}\,}

\def\Tr{\,\mbox{Tr}\,}
\def\sTr{\,\mbox{sTr}\,}

\def\al{\alpha}
\def\be{\beta}

\def\ga{\gamma}
\def\de{\delta}

\def\ep{\epsilon}

\def\na{\nabla}
\def\pa{\partial}

\def\ph{\varphi}

\def\th{\theta}

\def\Ga{\Gamma}

\begin{document}

\begin{center}
{\large\sc Quantum Effects in Softly Broken Gauge Theories  
\\
in Curved Space-Times}
\vskip 6mm
{\small \bf 
I.L. Buchbinder$^{a,b}$\footnote{E-mail address: joseph@tspu.edu.ru},
\quad 
G. de Berredo-Peixoto $^{b}$
\footnote{E-mail address: guilherme@fisica.ufjf.br},
\quad 
I.L. Shapiro$^{b}$}
\footnote{Also at Tomsk State Pedagogical University,
Russia. E-mail address: shapiro@fisica.ufjf.br} 
\vskip 6mm


{\small\sl
(a) \quad 
Department of Theoretical Physics, Tomsk State Pedagogical University
\\
634041, Tomsk, Russia
 
(b) \quad 
Departamento de F\'{\i}sica -- ICE,
Universidade Federal de Juiz de Fora} 
\\
{\small\sl  Juiz de Fora, CEP: 36036-330, MG,  Brazil }

\vskip 6mm
\end{center}

\vskip 4mm


\begin{quotation}
\noindent
{\large\bf Abstract.}
The soft breaking of gauge or other symmetries is the typical
Quantum Field Theory phenomenon. In many cases one can apply 
the St$\ddot{\rm u}$ckelberg procedure, which means introducing 
some additional field or fields and restore the gauge symmetry. 
The original softly broken theory corresponds to a particular 
choice of the gauge fixing condition. In this paper we use
this scheme for performing quantum calculations for some 
softly broken gauge theories in an arbitrary curved space-time. 
The following examples are treated in details: Proca field, 
massive QED and massive torsion coupled to fermion. Furthermore 
we present a qualitative discussions of the discontinuity of 
quantum effects in the massive spin-2 field theory, paying 
special attention to the similarity and differences with the 
massless limit in the spin-1 case. 
\vskip 1mm


\end{quotation}
\vskip 8mm

\section{\large\bf Introduction}

\quad
Theories with softly broken gauge symmetries emerge frequently in 
various branches of Quantum Field Theory (QFT). All attempts to 
consider the soft breaking of the nonabelian gauge symmetry met 
serious difficulties \cite{mYM}, but anyway they represent an 
important phase in the development of the modern high energy 
physics. One more example is supersymmetry which must be 
(most likely softly) broken in order to address the phenomenological 
applications and eventually experimental tests \cite{softSUSY}. 
Another interesting application of the soft symmetry breaking is 
the effective QFT approach to the propagating torsion
\cite{betor,guhesh,torsi}. The most relevant completely antisymmetric 
component of torsion can be described by the dual axial vector 
coupled to fermions through the axial vector current. The presence 
of the symmetry breaking mass of this axial vector is necessary for 
the consistency of the effective theory in the low-energy sector. 
Furthermore one can mention an important problem of discontinuity 
in the massless limit for the massive spin-2 (sometimes called 
massive graviton) field \cite{disc}. The massless and massive 
spin-2 particles have different number of degrees of freedom 
even if the mass is extremely small, hence there is no smooth 
massless limit, e.g., in the gravitational interaction. The problem 
of classical discontinuity can be solved if, instead of the flat 
background, one takes the curved one \cite{Kogan}, which in 
this case must be dS or AdS space. However, according to the 
publications \cite{duff-discont}, the discontinuity persists in the
quantum corrections, even in curved space. The last two examples 
show the importance of evaluating the quantum corrections in the 
theories with softly broken symmetries, especially in curved 
space-time. 

In the mentioned cases one is interested not only in the 
classical aspects of the theory, but also in deriving quantum 
corrections. The subject of the present paper is the calculation 
of effective action for softly broken gauge theories in curved 
space-time. In this case a kinetic term and interactions of 
classical action are gauge invariant while the massive terms 
are not. As a result the standard methods for evaluating 
effective action face serious technical difficulties. 
Our general strategy will be as follows: in each case 
we shall apply the St$\ddot{\rm u}$ckelberg procedure \cite{stuck},
that is, restore the gauge symmetry by introducing an extra field 
or a set of fields.  

The simplest example is the Proca field model in curved space, 
which is considered here as a kind of simple pedagogical example, 
illustrating the method. The restoration of gauge symmetry requires
introducing a new scalar field. Then, the original Proca theory 
corresponds to the special gauge fixing in a new theory, while
the quantum calculations are performed in some different gauge, 
which is most useful from technical 
viewpoint. Let us notice that the gauge fixing dependence of 
the effective action should vanish on shell. The situation is 
especially simple for the one-loop corrections, because in this 
case the difference between the effective actions calculated in 
different gauges is proportional to the classical equations of 
motion. Therefore, when evaluating quantum corrections to the 
vacuum action (that is the action of external, e.g., gravitational 
field), the result is gauge fixing independent. If we are dealing 
with the interacting theory and look also for the renormalization 
in the matter sector, some additional effort may be requested. 

As we shall see in what follows, our approach paves the way for much 
simpler and more efficient calculation of quantum corrections. The 
difference is especially explicit for the massive torsion-fermion 
system which was originally elaborated in \cite{guhesh}. The present 
method provides an independent verification of our previous result 
\cite{guhesh} and also enables one to perform the calculations in 
an arbitrary curved space-time, something that was impossible in the 
framework used in \cite{guhesh}. The paper is organized 
as follows. In section 2 we consider quantum calculations for the 
Proca field in curved space. In section 3 the result is generalized
for the massive QED and we also learn some important aspects of 
dealing with interacting fields. Section 4 is devoted to the 
massive torsion-fermion system. In section 5 we discuss the one-loop 
calculations for the massive spin-2 field, especially focusing on
the problem of discontinuity of quantum corrections in the massless 
limit. Finally, in section 6 we draw our conclusions.

\section{\large\bf Proca theory in curved space}

\quad
As a first example, consider the massive Abelian vector fields, which 
is also called Proca model. The action of the theory in curved space
has the form 
\beq
S_P = \int d^4x \sqrt{g}\,\Big\{
- \frac14\,F_{\mu\nu}^2 - \frac12\,M^2\,A_\mu^2\Big\}\,.
\label{Proca}
\eeq
Here and throughout the rest of the paper we use the Euclidean 
signature and condensed notations $F_{\mu\nu}^2=F_{\mu\nu}F^{\mu\nu}$ 
and $A_\mu^2=A_\mu A^\mu$. Also, we consistently disregard 
irrelevant surface terms. The main problem in deriving the quantum
corrections here is the softly broken gauge invariance. As a 
consequence of that, the bilinear form of the action 
\beq
{\hat H}=H_\al^\mu\,=\,
\de_\al^\mu \Box \,-\, \na_\al\na^\mu \,-\, R_\al^\mu - M^2\de_\al^\mu 
\label{Proca form}
\eeq
is degenerate while the theory is not invariant under the standard
gauge transformation. The non-invariance does not permit the use of 
the usual Faddeev-Popov technique for eliminating the degeneracy.
The known way of solving this problem \cite{bavi85} requires 
introducing an auxiliary operator
\beq
{\hat H}^*={H^*}_\mu^\nu = - \na_\mu \na^\nu + M^2\de_\mu^\nu \,,
\label{Proca 0}
\eeq
which satisfies the following two properties:
\beq
H_\al^\mu\,{H^*}_\mu^\nu &=& 
M^2\,\left(\de_\al^\nu \Box - R_\al^\mu - M^2\de_\al^\mu\right)\,,
\nonumber
\\
\Tr\ln {\hat H}^* &=& \Tr\ln \left(\Box - M^2\right)\,.
\label{Proca 1}
\eeq
As a result we arrive at the following relation\footnote{Indeed 
the one-loop effective action is given by the formula
\ $\bar{\Ga}^{(1)}=-\frac12\,\Tr\ln {\hat H}$ (see, e.g., 
\cite{bavi85,book}).}:
\beq
-\frac12\,\Tr\ln {\hat H} = 
-\frac12\,\Tr\ln 
\left(\de_\al^\nu \Box - R_\al^\mu - M^2\de_\al^\mu\right) 
+ \frac12\,\Tr\ln \left(\Box - M^2\right) \,.
\label{Proca 2}
\eeq
An obvious advantage of the last formula is that both operators at 
the {\it r.h.s} are not degenerate and admit a simple use of the 
standard Schwinger-DeWitt technique for the divergences and even 
the use of a more advanced method for deriving the non-local terms 
in the second-order in curvature approximation \cite{bavi90,fervi}
Looking at the expression (\ref{Proca 2}) one can observe certain 
similarity with the massless case. In both cases we meet 
contributions from minimal vector and scalar operators. Indeed, 
the second contribution in (\ref{Proca 2}) is analogous to the 
ghost contribution in the massless case, but there is a factor 
1 instead of a factor $1/2$ in front of the term $\,\Tr\ln \Box\,$
in the strictly massless case. As a result of this difference one 
can observe 
a discontinuity in the vacuum contribution of massive vector field 
in the massless limit. The difference between the $M\to 0$ limit 
in eq. (\ref{Proca 2}) and the contribution of a massless vector 
is exactly equal to the contribution of a minimal massless scalar. 
Which scalar is that? 

In order to understand that, let us consider a new action
\beq
S_P^\prime = \int d^4x \sqrt{g}\,\left\{
- \frac14\,F_{\mu\nu}^2 - \frac12\,M^2\,
\Big(A_\mu - \frac{1}{M}\,\pa_\mu \ph\Big)^2\right\}\,.
\label{Proca mod}
\eeq
The remarkable property of this expression is the gauge 
invariance under the simultaneous transformations 
\ $A_\mu \to A^\prime_\mu = A_\mu + \pa_\mu \xi$ \ and 
\ $\ph \to \ph^\prime = \ph + \xi M$. Furthermore, in the special 
gauge \ $\ph=0$ \ we come back to the Proca field action 
(\ref{Proca}). And finally, since both (\ref{Proca}) and 
(\ref{Proca mod}) are free fields actions, the gauge fixing 
dependence is irrelevant for the quantum correction which 
depends only on the external metric field. 

The last observation means it is not necessary to perform 
practical calculations in the in the $\ph=0$ gauge. Instead, 
one can choose another gauge, e.g., the one which simplifies 
the Feynman diagrams or the Schwinger-DeWitt technique. 
Let us use the linear gauge fixing condition
\ $\chi = \na_\mu A^\mu - M\ph$ \ for deriving the quantum 
corrections. Then the sum of the action (\ref{Proca mod})
and the FP gauge fixing term  \ 
$S_{gf}= - \frac12\,\int d^4x\sqrt{g}\,\chi^2$ \ is cast 
into the factorized form 
\beq
S^\prime + S_{gf} = \int d^4x \sqrt{g}\,\left\{ A^\al
\left(\de_\al^\nu \Box - R_\al^\mu - M^2\de_\al^\mu\right)A_\nu 
+ \ph\left(\Box - M^2\right)\ph \right\}\,.
\label{Proca 3}
\eeq
A simple calculation of the gauge ghost operator gives 
\beq
{\hat H}_{gh} = \Box - M^2
\label{Proca 4}
\eeq
and therefore the one-loop effective action is given by 
\beq
\bar{\Ga}^{(1)} \,=\, 
- \frac12\,\Tr\ln \left(\de_\al^\nu \Box 
- R_\al^\mu - M^2\de_\al^\mu\right) 
- \frac12\,\Tr\ln \left(\Box - M^2\right) 
+ \Tr\ln \left(\Box - M^2\right) \,,
\label{Proca 5}
\eeq
that is nothing but (\ref{Proca 2}). One can see that an 
extra scalar was indeed ``hidden'' in the massive term of the 
vector. At this point we conclude that the 
St$\ddot{\rm u}$ckelberg procedure described above works also 
in curved space-time and represents a useful alternative to the 
scheme (\ref{Proca 0})-(\ref{Proca 2}) for the Proca field in 
curved space-time \cite{bavi85}. 

Now we are in a position to discuss the discontinuity in the 
quantum contributions to the vacuum effective action from the 
Proca model in the massless limit. For this end we remember the 
second order in curvatures result for the Proca field derived in 
\cite{fervi}
\beq
{\bar \Ga}^{(1)}_{vector}
&=&\frac{1}{2(4\pi)^2}\,\int d^4x \,g^{1/2}\,
\Big\{\,\frac{3}{2}\,M^4\cdot\Big(\frac{1}{\ep}
+\frac32\Big)\,+\,\frac{M^2}{2}\,R\,\Big(\frac{1}{\ep}+1\,\Big)
\nonumber
\\
&+& \frac12\,C_{\mu\nu\al\be} \,\Big[\,\frac{13}{60\,\ep} + k^v_W(a)
\,\Big] C^{\mu\nu\al\be}
\,+\,R \,\Big[\,\frac{1}{72\,\ep}\,+\, k^v_R(a)\,\Big]
\,R\,\Big\}\,,
\label{final-v}
\eeq
where 
$$
\frac{1}{\ep}
= \frac{2}{4-n}+\ln \Big(\frac{4\pi \mu^2}{M^2}\Big)
- \gamma\,,
$$ 
$n$ \ is the parameter of dimensional regularization
and \ $\gamma$ \ is the Euler number. The nonlocal finite part 
of the effective action is characterized by the two formfactors
\beq
k^v_W(a) &=&
\,-\frac{91}{450}+\frac{2}{15a^2}
-\frac{8A}{3a^2}+A+\frac{8A}{5a^4}\,,
\nonumber
\\
k^v_R(a) &=& -\,\frac{1}{2160}+\frac{A}{48}
+\frac{A}{3a^4}+\frac{1}{36a^2}-\frac{A}{18a^2}\,.
\label{Cv}
\eeq
Here, $a$ and $A$ are defined according to 
\beq
A = 1-\frac{1}{a}\ln \Big|\frac{2+a}{2-a}\Big|
\quad {\rm and} \quad 
a^2=\frac{4\Box}{\Box+4M^2}\,.
\label{A}
\eeq

The detailed discussion of the massless limit in the formfactor 
$k^v_R(a)$ has been given in \cite{fervi} and also in \cite{AGS}
in relation to the ambiguity of the trace anomaly. Let us now pay 
attention to another formfactor $k^v_W(a)$. This term has very 
special importance because it defines the leading-log quantum 
contribution to the propagation of the gravitational wave, that 
is the transverse traceless part of the gravitational 
perturbation \ $h_{\mu\nu}=g_{\mu\nu}-g^{(0)}_{\mu\nu}$, where 
 \ $g^{(0)}_{\mu\nu}$ \ is the metric of background where the 
wave is propagating. We can take \ $g^{(0)}_{\mu\nu}=\eta_{\mu\nu}$
\ for simplicity. In the limit \ $M\to 0$ \ the expression \
$k^v_W(a)$ \ becomes
\beq
k^v_W(a)=\frac{13}{60}\,\Big(
\frac{1}{2-w}-\ga-\ln \frac{\Box}{4\pi \mu^2}\Big)-\frac{38}{225}\,.
\label{massless}
\eeq
The divergent term is cancelled by counterterm and the 
finite constant terms may be included into renormalization 
of the $C_{\mu\nu\al\be}^2$ term in the classical vacuum 
action. Finally, the most relevant term is of course the 
nonlocal one \ $-\frac{13}{60}\,\ln \frac{\Box}{4\pi \mu^2}$. 
This term is a physical contribution to the gravitational wave 
equation for the massless limit of the Proca model. On the other 
hand, the corresponding term derived for the gauge vector field 
is just \ $-\frac{1}{5}\,\ln \frac{\Box}{4\pi \mu^2}$. 
The difference between the two coefficients is
\  $1/60 = 13/60-1/5$ \ is nothing else but the contribution 
of an extra scalar field which we have discussed before. In the 
massless limit this field does not disappear and gives contribution 
to the vacuum effective action. This effect demonstrates the
discontinuity in the massless limit for the quantum contributions 
of the massive (Proca) vector field. 

\section{\large\bf Massive softly broken QED in curved space}

\quad
Our next example is the massive QED in curved space-time. 
The action has the form 
\beq
S \,=\, S_P \,+\,S_F\,,\quad
S_F \,=\, \frac{i}{2}\int d^4x\sqrt{g}\,\left\{\,
\bar{\psi}\ga^\mu D_\mu\psi - D^{\star}_\mu\bar{\psi}\ga^\mu\psi
+ 2i\bar{\psi}\, m\,\psi\,\right\}\,.
\label{action}
\eeq
where \ $D_\mu = \na_\mu + i\,q\,A_\mu\,;\quad 
D^{\star}_\mu = \na_\mu - i\,q\,A_\mu$, the operator $\na_\mu$ is the 
covariant derivative acting on Dirac fermion and \ $S_P$ \ has been 
defined in eq. (\ref{Proca}). The action (\ref{action}) possesses 
a softly broken gauge symmetry. The symmetry breaking term is the 
vector mass, so the situation is very similar to the one in the 
free Proca field case. There is an important difference, however. 
The theory (\ref{action}) includes interaction terms and those 
terms are gauge invariant. Therefore, in order to restore the 
gauge symmetry we should not just replace the vector field $A_\mu$ 
by the combination \ $A_\mu-\pa_\mu\ph/M$. At the same time, the 
procedure described in the previous section works perfectly well, 
so we now 
replace the $S_P$ in the action (\ref{action}) by the expression 
$S^\prime$ from (\ref{Proca mod}) and do not modify the term $S_F$ 
in (\ref{action}). It is easy to see that this procedure restores 
the symmetry and moreover the consideration of quantum corrections 
becomes very simple. 

In what follows we assume that the quantum corrections are derived 
within some explicitly covariant and gauge invariant approach, e.g.,
by using the 
background field method (see \cite{book} for the introduction).
Then in the vacuum 
(metric-dependent) sector one meets the quantum contributions 
described in section 2 plus the known contributions from the 
fermion \cite{fervi}, without additional discontinuity.  
Furthermore, in the matter sector the situation is rather simple 
too, at least at the one-loop level. The scalar sector is 
completely factorized, so the divergences of the theory are 
the same (except the vacuum ones, described in the previous 
section) as for the usual massive QED in curved space-time. 
The only problem which 
looks nontrivial is the relation between the divergences of the 
theory with softly broken symmetry (\ref{action}) and the 
divergences of the theory $S^\prime+S_F$ with restored gauge 
symmetry.  In order to address this question we remember that 
the theory (\ref{action}) corresponds to the particular gauge 
fixing condition $\ph=0$. On the other hand, 
the loop calculations are performed using the linear gauge fixing. 
What can be the difference between these two different gauges? 

In order to address the last question, let us remember three 
relevant facts. First, the divergences are local expressions. 
Second, they are gauge invariant. Third, the difference between
the divergent parts of one-loop Effective Actions obtained 
using two different gauges must be proportional to the 
classical equations of motion. In our case we denote 
${\bar \Ga}^{(1)}(\chi)$ the effective action corresponding to 
an arbitrary gauge fixing $\chi$ and ${\bar \Ga}^{(1)}(\chi_0)$ 
-- to some ``minimal'' gauge fixing $\chi_0$, e.g., the one which 
is most useful for practical calculations. Then, using the third 
statement from the list above we arrive at the formula
\beq
{\bar \Ga}^{(1)}(\chi_1) - {\bar \Ga}^{(1)}(\chi_2)=
\int \sqrt{g}\Big\{
f^{A}_\mu(\chi)\,\frac{\de S^\prime}{\de A_\mu}
+ f^{{\bar \psi}}(\chi)\,\frac{\de S^\prime}{\de {\bar \psi}}
+ \frac{\de S^\prime}{\de {\psi}}\,f^{\psi}(\chi)
+ f^{\ph}(\chi)\,\frac{\de S^\prime}{\de {\ph}}\Big\}\,,
\label{1}
\eeq
where $f^{A}_\mu(\chi),\,\,f^{{\bar \psi}}(\chi),\,\,f^{\psi}(\chi)
,\,\,f^{\ph}(\chi)$ are unknown local functions of the background 
(average) fields. 
By dimensional reasons and using covariance and Lorentz invariance 
arguments we arrive at the expressions involving only unknown 
dimensionless parameters which we can call \ $p_{1,2,3}$
\beq
f^{A}_\mu(\chi)=p_1{A}_\mu\,,\quad
f^{{\bar \psi}}=p_2{\bar \psi}\,,\quad
f^{\psi}(\chi)=p_2\psi\,,\quad
f^{\ph}(\chi)=p_3\ph\,.
\label{2}
\eeq
Then, after simple calculations one can obtain
\beq
{\bar \Ga}^{(1)}(\chi_1) - {\bar \Ga}^{(1)}(\chi_2)=
\int \sqrt{g}\Big\{-\frac{p_1}{2}\,F^2_{\al\be}
+ p_1M^2A_\al A^\al + (p_3-p_1){M}\,A^\al\pa_\al\ph
\nonumber
\\
-p_3(\na \ph)^2 
+ p_2 i {\bar \psi}\left(\gamma^\al\na_\al-im\right)\psi
+ (p_1+p_2)q{\bar \psi}\gamma^\al A_\al\psi\Big\}\,.
\label{3}
\eeq
Now, the requirement of gauge invariance tells us that 
$p_1=p_3=0$. Finally, the gauge fixing arbitrariness in the
theory under consideration reduces to the usual ambiguity 
in the renormalization of the fermion wave function. As a 
result one can safely perform one-loop calculations in the 
gauge invariant theory with the action 
$\,S^\prime \,=\, S^\prime_P \,+\,S_F\,$ and attribute the 
result to the theory with the softly broken symmetry 
(\ref{action}). The consideration similar to the one 
presented above can be applied to the much more complicated 
case of dynamical theory of torsion. We shall discuss this 
issue in the next section. 

\section{\large\bf 
Massive torsion coupled to fermion in curved space}

\quad
Consider the one-loop renormalization
in the coupled torsion-fermion system. This calculation plays 
an important role for the analysis of the possibility to have 
a consistent effective quantum field theory of dynamical torsion
\cite{guhesh,torsi}.

Torsion $\,T^\alpha_{\;\beta\gamma}\,$ is an independent (along 
with the metric) characteristic of a space-time manifold. 
It is defined by the relation (see, e.g., \cite{hehl,torsi} 
for introduction)
$$
{\Gamma}^\alpha_{\;\beta\gamma} -
{\Gamma}^\alpha_{\;\gamma\beta} =
T^\alpha_{\;\beta\gamma}\,.
$$
It proves useful to divide torsion into three irreducible 
components $\,T_{\mu},\,S_{\mu},\,q_{\alpha\beta\mu}\,$  
as follows:
\beq
T_{\alpha\beta\mu} =
\frac{1}{3} \left( T_{\beta}g_{\alpha\mu} -
T_{\mu}g_{\alpha\beta} \right)
- \frac{1}{6} \varepsilon_{\alpha\beta\mu\nu}
\,S^{\nu} + q_{\alpha\beta\mu}\,.
\label{t1}
\eeq
Interaction to the Dirac fermion is described by the action
\cite{bush85,book,torsi}
\beq
S_f \,=\, \int d^4x\sqrt{g}\,\left\{\,i\bar{\psi}\gamma^\mu 
\big( \na_\mu + i \eta_1\gamma^5S_\mu + i\eta_2T_\mu\big)\psi 
+ m\bar{\psi}\psi \right\}\,,
\label{t2}
\eeq
where $\eta_1,\,\eta_2$ \ are nonminimal parameters and 
\ $\na_\mu$ \ is Riemannian covariant derivative (without 
torsion). The minimal case corresponds to the action 
\beq
S_{min,f} \,=\, \frac{i}{2}\int d^4x\sqrt{g}\,\left\{\,
\bar{\psi}\ga^\mu {\tilde \na}_\mu\psi 
- {\tilde \na}^{\star}_\mu\bar{\psi}\ga^\mu\psi
\,-\, 2im\,\bar{\psi}\psi\,\right\}\,,
\label{minimal action}
\eeq
where \ ${\tilde \na}_\mu$ \ is the covariant derivative 
{\it with} torsion. 
Direct calculations show that this action corresponds to the
values of the parameters $\eta_1 = 1/8,\;\;\eta_2 = 0$ in 
the more general expression (\ref{t2}). The quantum theory 
meets serious difficulties for a fixed non-zero value of 
$\eta_1$ \cite{bush85}. 
Therefore, in what follows we consider $\eta_1$ as
an arbitrary parameter and keep $\eta_2=0$ for simplicity.
This is equivalent to requesting that the torsion tensor
is completely antisymmetric
$T_{\al\be\ga}= - \frac{1}{6} \varepsilon_{\alpha\beta\ga\mu}
\,S^{\mu}$. 

\subsection{General considerations}

\quad
The consistency conditions of an effective quantum 
fermion-torsion model requires unitarity and renormalizability 
in the low-energy sector. In this way we arrive at the 
unique possible form for the action of the theory \cite{betor}
(see also \cite{guhesh,torsi}), where we changed notation from 
$\eta_1$ to $\eta$
\beq
S_{tf} = \int d^4x \sqrt{g}\,\left\{ - \frac14\,S_{\mu\nu}^2
- \frac12\,M^2 S^2_\mu \,\,+\,\, i\bar{\psi}\gamma^\mu 
\big( \na_\mu + i \eta\gamma^5S_\mu\big)\psi 
+ m\bar{\psi}\psi \,\right\}\,.
\label{t5}
\eeq
Here $S_{\mu\nu}=\pa_\mu S_\nu - \pa_\nu S_\mu$ and $M$ is 
the torsion mass, which can not be zero because of the 
conditions listed above (see \cite{betor,torsi} for more 
details).

The kinetic and interaction terms of the theory (\ref{t5}) 
possesses the gauge symmetry
\beq
\psi' = \exp\left(i\eta\ga^5\be\right)\,\psi
\,,\quad
{\bar {\psi}}' = {\bar {\psi}}\,\exp\left(i\eta\ga^5\be\right)
\,,\quad
S_\mu ' = S_\mu - \pa_\mu\be
\,,\quad \mbox{where}\quad
\be=\be(x)\,.
\label{t55}
\eeq
Both massive terms are not invariant under these 
transformations and hence the symmetry is softly broken. In order 
to learn the implications of this fact we need to evaluate 
one-loop and two-loop quantum corrections \cite{guhesh}. The main 
practical difficulty comes from the one-loop divergences calculus. 
The method of derivation developed in \cite{guhesh} has been based
on the Boulware parametrization for a massive vector \cite{bou}. 
This approach is efficient but rather complicated. 
In particular, one has to develop a special kind of 
Schwinger-DeWitt expansion in the transverse vector space
and work out many universal traces \cite{bavi85}. Technically,
this is one of the most complicated one-loop calculations 
done so far. At the same time the physical output looks 
remarkable, in particular one can rule out the possibility of 
an independent light (that is potentially observable) torsion 
field using the quantum field theory arguments. Therefore, it 
would be nice to have an independent verification for the result 
of the mentioned 1-loop calculations.  
The St$\ddot{\rm u}$ckelberg procedure \cite{stuck}
opens the door for solving this problem.

Following the approach discussed in the previous sections, 
we introduce a new scalar field $\ph$ and restore the gauge 
symmetry in a following way:
\beq
S_{tf}^\prime 
&=& \int d^4x \sqrt{g}\, \Big\{ - \frac14\,S_{\mu\nu}^2
+ \frac12\,M^2\Big(S_\mu - \frac{\pa_\mu \ph}{M}\,\Big)^2
\nonumber
\\
&+&  i\bar{\psi}\gamma^\mu 
\big(\na_\mu + i \eta_1\gamma^5S_\mu\big)\psi 
\,\,+\,\, m \,\bar{\psi}\,
\exp \left(\frac{2i\eta\,\ga^5\ph}{M}\right)\,\psi 
\Big\}\,,
\label{t555}
\eeq
The symmetry transformations (\ref{t55}) must be supplemented 
by \ $\ph \to \ph^\prime = \ph - M\be$. The original theory 
(\ref{t5}) is restored when we use the gauge fixing condition 
$\ph=0$. However, the practical calculations can be performed 
in some useful linear gauge where the scalar field persists. 
In this case we meet the new exponential coupling with the 
negative dimensional coupling. This is a very remarkable  
occurrence, because it provides a qualitative explanation 
for the result obtained in \cite{guhesh}. Because of the 
exponential coupling the theory (\ref{t555}) is not 
renormalizable and therefore it is not a surprise that we 
meet the counterterms with the form which is different from 
the ones in the classical action. In particular, one can
easily construct the divergent diagrams which 
produce the $\frac{1}{M^2}(\bar{\psi}\,{\psi})^2$-type
divergence. Indeed, we really met these divergences 
in \cite{guhesh} after very complicated calculations. Now
we can understand them as manifestation of the loop made 
from an extra scalar degree of freedom, which was hidden 
in the original formulation of the theory and became explicit 
after we used the St$\ddot{\rm u}$ckelberg procedure. 

Indeed, the eq. (\ref{t555}) is an essential generalization 
of the usual St$\ddot{\rm u}$ckelberg procedure \cite{stuck}
for the case of complicated theory. In the next subsection 
we demonstrate how this new transformation may be applied 
for quantum calculations. 

Before starting the one-loop calculation, let us briefly discuss 
the gauge fixing dependence in the theory (\ref{t555}). The 
expression for the difference between two effective actions 
corresponding to the different gauge fixing conditions is very 
similar to the one for the massive QED 
(\ref{3})
\beq
{\bar \Ga}^{(1)}(\chi_1) - {\bar \Ga}^{(1)}(\chi_2)=
\int \sqrt{g}\Big\{-\frac{p_1}{2}\,S^2_{\al\be}
+ p_1M^2S_\al S^\al + (p_3-p_1){M}\,S^\al\pa_\al\ph
-p_3(\na \ph)^2 
\nonumber
\\
+ p_2 i {\bar \psi}\left(\gamma^\al\na_\al-im\right)\psi
+ (p_1+p_2)q{\bar \psi}\gamma^\al S_\al\psi
\,+\, \frac{2im}{M}\,p_3\,\ph\,{\bar \psi} \,
\exp \left(\frac{2\eta\,\ga^5\ph}{M}\right)\,\psi\Big\}\,.
\label{3S}
\eeq
As in the previous case, the gauge invariance of the divergences
requires $p_1=p_3=0$ and does not impose restrictions on the 
parameter $p_2$. The gauge fixing dependence is reducing to the
standard ambiguity in the renormalization of the fermion wave 
function. 

\subsection{One-loop calculation}

\quad
In this subsection we shall calculate the one-loop divergences
in the theory (\ref{t555}), using the background field method
and Schwinger-DeWitt technique of extracting divergences of 
the Effective Action. After that we shall fix the gauge 
$\,\ph=0\,$ for the background fields and compare the result 
with the one obtained in \cite{guhesh} using much more 
complicated approach. The application of the 
St$\ddot{\rm u}$ckelberg procedure leads to serious improvement
in the calculational procedure. In particular, here we will be
able to obtain the divergences in curved space-time and hence 
our results will be more general than the ones of \cite{guhesh}.

The calculation will be performed using background field method
and standard (different from \cite{guhesh}) Schwinger-DeWitt 
technique (see, e.g., \cite{book} for the introduction). The 
first step is the splitting of the fields into background and 
quantum ones
\beq
S_\mu \to S_\mu + s_\mu\,,\quad
\psi \to \psi + \chi\,,\quad
\bar{\psi} \to  \bar{\psi} + \bar{\chi}\,,\quad
\ph \to \ph + \th\,,
\label{tob}
\eeq
where we kept classical notations for the background fields. 
The most simple ``minimal'' gauge-fixing term has the form
\beq
{\cal L}_{{\rm GF}} = -\frac{1}{2}\,\sqrt{g}\,
\left(\,\na_\mu s^\mu + M\th\,\right)^2\,.
\label{gauge t}
\eeq
The bilinear Lagrangian in quantum fields, including the 
gauge-fixing term, is given by
\beq
L_t^{(2)}=\frac12\,\left(s_\mu\,\,\bar{\chi}\,\,\th\right)
\,\,\big(\hat{{\cal H}}\big)\,\,
\left(\begin{array}{c}
s^\nu 
\\
{\chi}
\\
\th 
\end{array}
\right)\,,
\label{to1}
\eeq
where
\beq
\hat{{\cal H}} = \left(
\begin{array}{ccc}
\de^\mu_\nu \Box - R^{\mu}_{\nu} + M^2 \de^\mu_\nu & 
2\eta\bar{\psi}\ga^5\ga^\mu & 0 \\
 & & \\
2\eta \ga^5\ga_\nu \psi & 
\begin{array}{c} 
2 i \ga^\mu\nabla_\mu 
+ 2\eta\ga^5\ga^\mu S_\mu 
\cr
+ 2 m {\rm e}^{2i\eta\varphi\ga^5/M}
\end{array}
& \frac{4i\eta m}{M}\ga^5 {\rm e}^{2i\eta\varphi\ga^5/M}\psi \\
 & & \\
0 & \frac{4i\eta m}{M}\bar{\psi}\ga^5 
{\rm e}^{2i\eta\varphi\ga^5/M} &
\begin{array}{c}
 -\Box - M^2 
\cr
- \frac{4\eta^2 m}{M^2}\bar{\psi}{\rm e}^{2i\eta\varphi\ga^5/M}\psi  
\end{array}
\end{array} \right)\,.
\label{to2}
\eeq
Multiplying $\hat{{\cal H}}$ by a matrix $\hat{K}$
\beq
\hat{K} = \left(
\begin{array}{ccc}
\de^\nu_\al & 0 & 0 \\
0 & -\frac12 i\ga^\rho\na_\rho & 0 \\
0 & 0 & -1
\end{array} 
\right)\, ,
\eeq
we obtain the operator 
\beq
\hat{{\cal H}} \hat{K} = \hat{1}\Box + 
2\hat{h}^\rho\na_\rho + \hat{\Pi}\, ,
\eeq
where
\beq
\hat{1} = \left(
\begin{array}{ccc}
\de^\mu_\nu & 0 & 0 \\
0 & \hat{1} & 0 \\
0 & 0 & 1
\end{array}
\right)\,,\qquad 
\hat{h}^\rho = \left(
\begin{array}{ccc}
0 & -\frac{i}{2} \eta\bar{\psi}\ga^5\ga^\mu\ga^\rho & 0 \\
0 & -\frac{i}{2} m {\rm e}^{2i\eta\varphi\ga^5/M}\ga^\rho - 
\frac{i}{2} \eta\ga^5\ga^\al\ga^\rho S_\al & 0 \\
0 & \frac{\eta m}{M}\bar{\psi}\ga^5 
{\rm e}^{2i\eta\varphi\ga^5/M}\ga^\rho & 0
\end{array}
\right)
\eeq
and
\beq
\hat{\Pi} = \left(
\begin{array}{ccc}
-R^\mu_\nu + M^2\de^\mu_\nu & 0 & 0 \\
2\eta\ga^5\ga_\nu\psi & -\frac{\hat{1}}{4}R & 
-\frac{4i\eta m}{M}\ga^5 {\rm e}^{2i\eta\varphi\ga^5/M}\psi \\
0 & 0 & M^2 + \frac{4\eta^2 m}{M^2}\bar{\psi} 
{\rm e}^{2i\eta\varphi\ga^5/M} \psi 
\end{array}
\right)\, .
\eeq

The divergent part of \ $\sTr\ln\hat{{\cal H}}$ \ (here \ $\sTr$ \
means functional supertrace, including covariant integration over 
the spacetime variables, usual trace over spinor indices and taking 
into account the Grassmann parity of the fields) can be achieved 
just by calculating $\sTr\ln (\hat{{\cal H}} \hat{K})$ and then 
subtracting the $\sTr\ln \hat{K}$
\beq
-\frac{1}{2}\sTr\ln \hat{{\cal H}} 
= - \frac{1}{2}\sTr\ln (\hat{{\cal H}}\hat{K}) 
+ \frac{1}{2}\sTr\ln \hat{K} \,,
\eeq
where $\sTr\ln \hat{K}$ produces only metric-dependent vacuum 
contributions to the Effective Action. They are well known and 
in fact irrelevant for our purposes. In what follows we will 
not consider these terms. As far as the integration over the 
Faddeev-Popov (FP) ghosts also gives only vacuum (that is depending 
exclusively on the metric) contributions, one can find the 
relevant one-loop divergences in the form
\beq
\Gamma^{(1)}_{{\rm div}} = \left.
\frac{i}{2}\sTr\ln (\hat{{\cal H}}\hat{K})\right|_{{\rm div}}\, .
\eeq
The above expression can be calculated by using the standard
Schwinger-DeWitt algorithm
\beq
-\frac{1}{2}\sTr\ln (\hat{{\cal H}}\hat{K})|_{{\rm div}} = 
-\frac{\mu^{n-4}}{(4\pi)^2(n-4)}
\sTr \left\{ \frac{1}{2}\,\hat{P}\hat{P} + 
\frac{1}{12}\,\hat{{\cal R}}_{\al\be}\hat{{\cal R}}^{\al\be}
\right\}\, ,
\eeq
where the vacuum terms were omitted
and the matrices $\hat{P}$ and
$\hat{{\cal R}}_{\al\be}$ are given by
\beq
\hat{P} &=& \hat{\Pi}+\frac{\hat{1}}{6} R 
- \na_\rho \hat{h}^\rho - \hat{h}_\rho \hat{h}^\rho\,,
\nonumber
\\
\hat{{\cal R}}_{\al\be} &=& \big[\na_\be\,,\na_\al\big]\,\hat{1} +
\na_\be \hat{h}_\al - \na_\al \hat{h}_\be + \hat{h}_\be \hat{h}_\al
- \hat{h}_\al \hat{h}_\be\, . 
\nonumber
\eeq
The straightforward calculation using this formula gives us,
after certain algebra, the following result for the relevant 
terms of the Effective Action:
\beq
\Gamma^{(1)}_{{\rm div}} & = &
-\,\frac{\mu^{n-4}}{(4\pi)^2(n-4)} \int d^nx\sqrt{g}\,
\left\{
4\eta^2 m^2 S^\mu\Big(S_\mu - \frac{1}{M}\na_\mu\varphi\Big)
- \frac13\,\eta^2 S_{\mu\nu}^2 
+ 2i\eta^2\bar{\psi}\ga^\rho {\cal D}_\rho \psi  \right. 
\nonumber 
\\
& + & \left. 
\frac{4i\,\eta^2 m^2}{M^2}\,
\bar{\psi}\ga^\rho{\cal D}^{\ast}_\rho\psi 
+ 8\eta^2 m \Big(\frac{m^2}{M^2} - \frac12\Big)\bar{\psi}\,
\exp\Big(\frac{2i\eta\varphi\ga^5}{M}\Big) \,\psi\right. 
\\
& - & \left.
\frac{8\eta^3 m^2}{M^3}\bar{\psi}\ga^\mu\ga^5(\na_\mu\varphi)\,\psi + 
\frac{2\eta^2 m}{3M^2}\bar{\psi}{\rm e}^{2i\eta\varphi\ga^5/M}R\psi
+\frac{8\eta^4 m^2}{M^4}
\Big(\bar{\psi}{\rm e}^{2i\eta\varphi\ga^5/M}\psi\Big)^2 
\right\}\, ,
\nonumber
\eeq
where ${\cal D}_\rho = \na_\rho + i\eta\ga^5 S_\rho$ and 
${\cal D}^{\ast}_\rho = \na_\rho - i\eta\ga^5 S_\rho$. Let us
notice that this expression represents the divergences of the 
theory  (\ref{t555}), involving axial vector, fermion and 
scalar, with exponential interaction between the last two 
fields. This result has its own independent merit, especially 
because the interaction is non-polynomial and the above 
expression is a sum of an infinite number of Feynman 
diagrams. 

In order to obtain the one-loop divergences for the original 
theory (\ref{t5}), one has to put \ $\varphi = 0$.
Then the above result reduces to 
\beq
\Gamma^{(1)}_{{\rm div}} & = &
-\frac{\mu^{n-4}}{(4\pi)^2(n-4)} \int d^nx\sqrt{g}\left\{
4\eta^2 m^2 S^\mu S_\mu - \frac13 \eta^2 S_{\mu\nu}^2 
+ 4i\eta^2\frac{m^2}{M^2}\bar{\psi}\ga^\mu {\cal D}_\mu^{\ast}\psi
+ \right. \nonumber \\
& + & \left.
2i\eta^2\,{\bar \psi}\ga^\mu {\cal D}_\mu\psi
+ \left(\frac{8\eta^2 m^3}{M^2} - 4\eta^2 m \right)\bar{\psi}\psi
+ \frac{2\eta^2 m}{3M^2}\bar{\psi}\,R\,\psi 
+ \frac{8\eta^4 m^2}{M^4}(\bar{\psi}\psi)^2  
\right\}\,.
\label{r1}
\eeq
It is easy to see that the difference with the one-loop result 
derived by different method in \cite{guhesh} consists in the 
following two kind of terms:

1) Term proportional to the scalar curvature could not be 
obtained in \cite{guhesh}, because the calculation has been 
performed in flat space-time. 

2) For the flat background, the difference between the two
results
$$
\Gamma\{\mbox{Eq. (\ref{r1})}\} 
\,-\, \Gamma \{\mbox{Ref. \cite{guhesh}}\}
\quad \propto \quad
2\eta^2\,\bar{\psi}\,\left(\,i\ga^\mu {\cal D}_\mu\psi 
+ m\,\right) \psi
$$
is proportional to the classical equations of motion
for the fields and hence, according to the considerations 
presented in the previous section, this difference is due 
to the distinct parametrizations of the quantum fields in 
two cases. 
All in all, the expression (\ref{r1}) represents a perfect 
fit to the result of the much more involved and cumbersome 
calculation of \cite{guhesh}.
 

\section{\large\bf Qualitative discussion of a discontinuity 
phenomena in massive spin-1 and spin-2 field theories.}

\quad
Free massive higher spin field models (see, e.g., 
\cite{SH,DN,H,B,BGKP,Z,BLK}) are typical examples of the 
softly broken gauge theories, where the gauge symmetry is 
broken by the non-zero mass of the field. The
evaluation of effective action in these theories can be carried 
out with the help of St$\ddot{\rm u}$ckelberg procedure. 
This procedure has been used for the effective action in the 
massive spin-1 field model in curved space in Section 2 and in 
the massive spin-2 field model on AdS space in \cite{duff-discont}. 
We do not intend to perform practical 
calculations for higher spin fields in the present paper. Instead 
we shall address one important general aspect of the quantum 
corrections produced by higher spin fields on curved background. 
For this end we perform a comparison of the discontinuity phenomena 
for the massive theories with spin-1 and spin-2.

In the massless limit, in flat space-time, the massive spin-2 field
manifests a discontinuity \cite{disc} due to the different number
of degrees of freedom in the massive and massless cases. The
situation may be quite different in the curved space. In particular,
it has been shown that the classical discontinuity does not occur in
the Anti-de Sitter \cite{H} and in the de Sitter \cite{Kogan} space.
At the same time, according to \cite{duff-discont} the discontinuity
persists in the vacuum quantum corrections generated by the massive
spin-2 field.

The calculations performed in \cite{duff-discont} are quite
similar to the ones we have presented in section 2 for the massive
vector field. The St$\ddot{\rm u}$ckelberg procedure requires the
introduction of vector and scalar auxiliary fields, which are called
to restore the symmetry which is softly broken by massive fields. In
the massless limit \ $m^2 \to 0$ \ the loops of the auxiliary
fields do not disappear and moreover their number is not the same as
the number of the Faddeev-Popov ghosts in the massless diffeomorphism
invariant case. 

As a result the divergent parts of the vacuum effective actions  
are indeed different for the massless spin-2 field and for the 
massless limit of the massive spin-2 field. At the first sight, 
the relation between quantum corrections for $m\to 0$ and $m=0$ 
cases in the \ spin-2 theory looks similar to the one in the vector 
case. 

One can notice that there is also a significant difference
between the discontinuity of the quantum corrections in the
massive spin-1 and spin-2 cases. Let us look again at the
expression for the \ $m\to 0$ \ limit of the quantum
contribution for the massive spin-1 field (\ref{massless}).
As it was already discussed in section 2, we can distinguish
two different parts in this expression. The first is local,
it includes the divergent expression. One has to remember
that this part has no direct physical sense, for it must be
modified by adding a local counterterm in the vacuum sector
\cite{book}. After that, the overall coefficient of the
local term should be fixed by the renormalization condition.
Then, the difference between the theories with different
coefficients of the {\it local terms} disappears. Of course,
the last statement is not really correct in the vector
case, where we also meet a very important non-local term, with
the coefficient which is equal to the one of the local divergent
term. This equality is important, in particular it provides
the possibility to derive the renormalization group
$\be$-functions in the \ $\overline{\rm MS}$ \ renormalization
scheme.

However, a Lagrangian construction for the free massive spin-2 field
model is known only for the spaces of constant curvature (see, e.g.,
\cite{DN,H,B,BGKP,Z,BLK}). In such
spaces one can not construct the non-local insertions similar to 
the ones we met in eq. (\ref{massless}). The physical sense of the
renormalization group in the vacuum sector becomes, in this case,
rather obscure. As a result, all the existing difference between the
massless theory and the massless limit of the massive theory can be
formally eliminated by the renormalization of vacuum effective action. 
In spite of the difference between the divergent contributions 
to vacuum effective action in the two cases ($m=0$ {\it vs} $m\to 0$) 
can be treated as manifestation of discontinuity, it does not 
automatically imply that there is a discontinuity in the quantum 
corrections to physically observable quantities. 
In order to clarify this problem one has to achieve the physical 
interpretation of the available difference between quantum corrections 
to renormalized quantities for the cases of zero and non-zero masses
in the spin-2 filed theories.

\section{\large\bf Conclusions}

\quad
We have demonstrated the advantages of using the St$\ddot{\rm
u}$ckelberg procedure for the purposes of quantum calculations in
curved space-time. In the case of massive Abelian vector we have
calculated, in a new alternative manner, the one-loop correction to
the graviton propagator and investigated the discontinuity of this
correction in the massless limit. Furthermore, we applied the same
procedure to the two different models of interacting spin-1 and
fermion field. In the case of massive QED the one-loop calculation
is very simple and we just arrive at the result which can be, in
principle, obtained by other methods.

In the case of axial vector (antisymmetric torsion) coupled to
fermion we arrived at the new method of calculation which is much
better than the previously developed one \cite{guhesh} in many
respects. In particular, the calculations become much simpler and
they could be completed even in an arbitrary curved background. It
is remarkable that in the flat space limit we have confirmed the
previous result, obtained by in a much more cumbersome way. The new
approach based on the St$\ddot{\rm u}$ckelberg procedure for the
axial vector, made the statement about the non-renormalizability of
the theory much more explicit and, once again, demonstrated an
essential similarity with the massive Yang-Mills case \cite{mYM}
(see also recent works \cite{KMKN}).

Finally, we presented a qualitative discussion of a very interesting
problem of discontinuity of the massless limit for the massive
spin-2 contributions to the renormalization of the vacuum energy.
The practical derivation of the one-loop contributions of the spin-2
fields has been performed in \cite{duff-discont} by using the
St$\ddot{\rm u}$ckelberg procedure. We have argued, using comparison
with the spin-1 case that the interpretation of this quantum
discontinuity is not obvious at the present state of knowledge about
the massive higher spin fields on curved background. At the same
time, the calculations of effective action using the St$\ddot{\rm
u}$ckelberg procedure can be performed for the massive fields of an
arbitrary spins on AdS space. The expected result is qualitatively
similar with ones for the $s=2$ case, however the algebraic
structure of the relevant operators may be a bit more involved. Of
course, a problem of effective action for massive arbitrary higher
spin theories on AdS space is interesting and important itself. We
plan to report about the practical calculations in the massive $s=3$
and maybe other cases in the near future.
\vskip 6mm

\noindent
{\large\bf Acknowledgments.}
I.L.B is grateful to FAPEMIG (Minas Gerais, Brazil) for the finance 
support of his visit to the Physics Department of UFJF. The work of 
I.L.B has been partially supported by the grant for LRSS, project 
No 4489.2006.2, joint RFBR-DFG grant, project No 06-02-04012,  
DFG grant, project No 436 RUS/13/669/0-3 and INTAS grant,  
project INTAS -05-7928.  G.B.P. and I.Sh acknowledge support of CNPq 
(Brazil), FAPEMIG, FAPES and ICTP.
\vskip 6mm


\end{document}